\documentclass[aps,twocolumn,showpacs,preprintnumbers,amsmath,amssymb,superscriptaddress,floatfix,nofootinbib]{revtex4}
\usepackage{lipsum}
\usepackage{graphicx}
\usepackage{epsfig}
\usepackage{epstopdf}
\usepackage{hyperref}
\usepackage{amsmath}
\usepackage{amsfonts}
\usepackage{amssymb}

\begin{document}

\title{$f_0(500)$, $f_0(980)$ and $a_0(980)$ production in the $\chi_{c1} \to \eta \pi^+\pi^-$ reaction}

\author{Wei-Hong Liang}
\email{liangwh@gxnu.edu.cn}
\affiliation{Department of Physics, Guangxi Normal University,
Guilin 541004, China}

\author{Ju-Jun Xie}
\email{xiejujun@impcas.ac.cn}
\affiliation{Institute of Modern Physics, Chinese Academy of
Sciences, Lanzhou 730000, China} \affiliation{Research Center for
Hadron and CSR Physics, Institute of Modern Physics of CAS and
Lanzhou University, Lanzhou 730000, China}

\author{E.~Oset}
\email{oset@ific.uv.es}
\affiliation{Institute of Modern Physics, Chinese Academy of
Sciences, Lanzhou 730000, China} \affiliation{Departamento de
F\'{\i}sica Te\'orica and IFIC, Centro Mixto Universidad de
Valencia-CSIC Institutos de Investigaci\'on de Paterna, Aptdo.
22085, 46071 Valencia, Spain}

\date{\today}

\begin{abstract}
We study the $\chi_{c1} \to \eta \pi^+ \pi^-$ decay, paying attention
to the production of $f_0(500)$, $f_0(980)$ and $a_0(980)$ from the
final state interaction of pairs of mesons that can lead to these
three mesons in the final state, which is implemented using the chiral
unitary approach. Very clean and strong signals are obtained for the
$a_0(980)$ excitation in the $\eta \pi$ invariant mass distribution and
for the $f_0(500)$ in the $\pi^+ \pi^-$ mass distribution. A smaller,
but also clear signal for the $f_0(980)$ excitation is obtained. The
results are contrasted with experimental data and the agreement found is
good, providing yet one more test in support of the picture where these
resonances are dynamically generated from the meson-meson interaction.
\end{abstract}

\maketitle

\section{Introduction}

The $\chi_{c1} \to \eta \pi^+ \pi^-$ reaction has been measured by
the CLEO collaboration in Ref. \cite{Adams:2011sq} and is presented as
the reaction where a cleanest signal for the $a_0(980)$ resonance is
seen. Indeed, a neat and strong peak is observed in the $\eta \pi$
invariant mass distribution, peaking around the $K \bar K$ threshold
and with the characteristic strong cusp structure of this resonance,
as observed in other high statistics experiments \cite{rubin}. What
makes this experiment singular is that the strength of the peak is
much bigger than the rest of the distribution at other  $\eta \pi$
invariant masses. The complementary $\pi^+ \pi^-$ mass distribution
shows a clear contribution from the $f_0(500)$ resonance at lower
invariant masses, a dip in the region of the  $f_0(980)$ and also a
strong peak for the $f_2(1270)$ resonance and of the $f_4(2050)$ at
larger invariant masses. The reaction has been remeasured with much
more statistics by the BESIII collaboration and is presently under
internal discussion. A preliminary view of the results is available
in Ref. \cite{kornicer}, where in the region of the $f_0(980)$ a small
peak seems to show up followed by a dip around 1070 MeV. This hence constitutes a
clear case for a test of the ideas of the unitarized chiral
perturbation theory, the chiral unitary approach. In this approach
the input from chiral Lagrangians \cite{gasser} for the meson-meson
interaction is used in a coupled channels Bethe Salpeter equation,
from where the $f_0(500)$,  $f_0(980)$ and $a_0(980)$ resonances
emerge \cite{npa,kaiser,markushin,juanito}. They are dynamically
generated from the meson-meson interaction and would qualify as
meson-meson molecular states. The same results are obtained using an
equivalent unitarizing method, the inverse amplitude method, in
Refs. \cite{ramonet,rios}.

The nature of the low energy scalar mesons has generated a long
controversy \cite{Klempt:2007cp}.  Yet, other different approaches
which start from a seed of $q \bar q$ for these resonances, also get
a large meson-meson component for these states as soon as this seed
is coupled to two mesons and the mesons are allowed to interact in a
realistic scheme fulfilling unitarity
\cite{vanBeveren:1986ea,Tornqvist:1995ay,Fariborz:2009cq,Fariborz:2009wf}.
A thorough recent review on these issues can be found in Ref. \cite{sigma},
presenting theoretical arguments and abundant experimental
information that support the picture of the dynamical generation of
these resonances and its clear difference from $q \bar q$ states.

The chiral unitary approach not only provides a picture for these
resonances, it also allows to make clean predictions for any
reaction where these resonances are produced, providing, in the
worse of the cases, when not enough dynamical information is
available for the process studied, ratios for the production of the
different resonances. This is a remarkable property of this approach
that is not shared by other theoretical approaches trying to
interpret the data. Hence experimental data could easily disprove
the model, but so far this has not been the case in spite of the many
reactions studied (see a recent review of $B$ and $D$ decays where
many such reactions are analyzed and discussed \cite{weakreview}).
Two of the most recent cases are the $B^0$ and $B^0_s$ decays into
$J/\psi \pi^+ \pi^-$ measured in Ref. \cite{Aaij:2011fx} and analyzed in
Ref. \cite{liang} (see Refs. \cite{Wang:2015uea,hanhart} for a different
approach based on the use of form factors) and the $D^0$ decay into
$K^0$ and the $f_0(500)$, $f_0(980)$ and $a_0(980)$ measured in
Ref. \cite{Muramatsu:2002jp} and analyzed in Ref. \cite{dai} (see
Ref. \cite{robert} for also an approach based on form factors). Yet, the
present reaction, with its spectacular signal for $a_0(980)$
production, a large signal for $f_0(500)$ and the small signal for
the $f_0(980)$, all seen in the same reaction, is a case that should
not be missed to challenge this theoretical approach. The purpose of
the present paper is to make the theoretical study of the process
along the lines of the chiral unitary approach, to confront the
results with the relevant data already existing and eventually
predict some features that could also be detected with the coming
analysis from the BESIII large statistics experiment.

\section{Formalism}
\begin{figure}[tb]\centering
\includegraphics[scale=0.45]{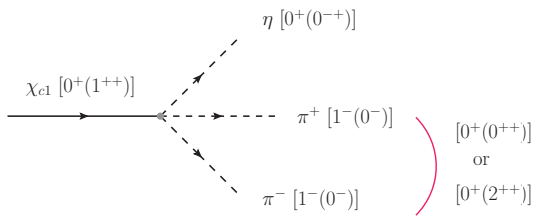}
\caption{$\chi_{c1} \to \eta \pi^+ \pi^-$ process with the quantum numbers of the particles produced. The $\pi^+ \pi^-$ pair combines to $f_0(500), f_0(980)$ and $f_2(1270)$.
\label{fig:FeynmanDiag1}}
\end{figure}

The $\chi_{c1} \to \eta \pi^+ \pi^-$ decay is depicted in Fig. \ref{fig:FeynmanDiag1}
with the quantum numbers of the different particles.
The  $\chi_{c1}$ has $I^G(J^{PC})\equiv 0^+(1^{++})$
and the $\eta$, $0^+(0^{-+})$. The conservation of quantum numbers indicates that the $\pi^+ \pi^-$ pair must have isospin $I=0$, $C$-parity positive and $G$-parity positive. In addition, since the $\chi_{c1}$ has spin 1 we need one unity of spin or angular momentum in the final state. Since neither of the final $\eta$, $\pi^+$, $\pi^-$ has spin, we need to form a scalar with the polarization vector of the $\chi_{c1}$ and a momentum of one of the mesons. We can have a structure like
\begin{align}\label{eq:V_structure}
V_1 &= A \vec{\epsilon}_{\chi_{c1}} \cdot \vec{p}_{\eta}, \nonumber \\
V_2 &= B \vec{\epsilon}_{\chi_{c1}} \cdot \vec{p}_{\pi^{+}}, \\
V_3 &= C \vec{\epsilon}_{\chi_{c1}} \cdot \vec{p}_{\pi^{-}}. \nonumber
\end{align}
Let us take the first structure of $V_1$. The $\vec{p}_{\eta}$ coupling introduces $L=1$ and forces the $\pi^+ \pi^-$ pair to also have positive parity. With these quantum numbers, the $\pi^+ \pi^-$ pair can be $0^+ (0^{++})$, $0^+ (2^{++})$, and then can produce the resonance $f_0(500)$, $f_0(980)$ and $f_2(1270)$, which are well known resonances.

\begin{figure}[tb]\centering
\includegraphics[scale=0.45]{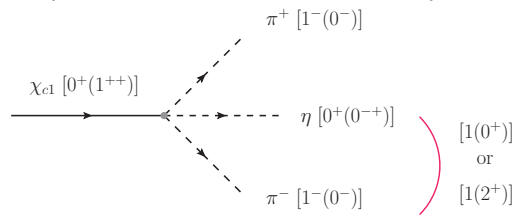}
\caption{$\chi_{c1} \to \eta \pi^+ \pi^-$ process with the quantum numbers of the particles produced. The $\pi^- \eta$ pair combines to $a_0(980)$ and $a_2(1320)$.
\label{fig:FeynmanDiag2}}
\end{figure}

Let us single out the term of $V_2$ in Eq. (\ref{eq:V_structure}). Now it is the $\pi^+$ the one that carries $L=1$. We write for this term another diagram in Fig. \ref{fig:FeynmanDiag2}.
Now the $\eta \pi^-$ system must have $I=1$ and positive parity. The angular momentum of $\pi^- \eta$ can be $L'= 0,2$ and then we can have as primary choice $a_0(980)$ production and in the analysis of Ref. \cite{Adams:2011sq} they also allow $a_2(1320)$ formation.

As we can see, we can produce $f_0(500)$, $f_0(980)$ and $a_0(980)$ in the same reaction and there is still one more symmetry that we must consider and which, together with the ingredients of the chiral unitary approach, will allow us to establish the connection between the production of any of them in this reaction. The symmetry that we invoke is SU(3) symmetry.
Since the $\chi_{c1}$ is a $c\bar c$ state, with respect to the $u,d,s$ quarks, it behaves as a neutral system, it is a scalar of SU(3). Thus, we must construct a scalar of SU(3) with three pseudoscalar mesons. This means that we will inevitably mix the $\eta \pi^+ \pi^-$ with other three meson states that can appear in the $\chi_{c1}$ decay. This will occur at a primary step of the $\chi_{c1}$ decay, but then the mesons will interact in coupled channels and finally produce the $\eta \pi^+ \pi^-$ in a final step.

In order to see the proper combination of three mesons that lead to a SU(3) scalar, we introduce the $q\bar q$ matrix $M$
\begin{equation}\label{eq:M_matrix}
M=\left(
           \begin{array}{ccc}
             u\bar u & u \bar d & u\bar s \\
             d\bar u & d\bar d & d\bar s \\
             s\bar u & s\bar d & s\bar s \\
           \end{array}
         \right)=  \left(\begin{array}{c}
u \\ d \\ s
\end{array}\right)
\left(\begin{array}{ccc}
\bar{u} & \bar{d} & \bar{s}
\end{array}\right).
\end{equation}
This matrix has the property
\begin{align}\label{eq:M_matrix-2}
&MMM  \nonumber\\
&=
\left(\begin{array}{c}
u \\ d \\ s
\end{array}\right)
\left(\begin{array}{ccc}
\bar{u} & \bar{d} & \bar{s}
\end{array}\right)
\left(\begin{array}{c}
u \\ d \\ s
\end{array}\right)
\left(\begin{array}{ccc}
\bar{u} & \bar{d} & \bar{s}
\end{array}\right)
\left(\begin{array}{c}
u \\ d \\ s
\end{array}\right)
\left(\begin{array}{ccc}
\bar{u} & \bar{d} & \bar{s}
\end{array}\right)  \nonumber\\
&=
\left(\begin{array}{c}
u \\ d \\ s
\end{array}\right)
\left(\begin{array}{ccc}
\bar{u} & \bar{d} & \bar{s}
\end{array}\right) (\bar u u +\bar d d +\bar s s)^2 \nonumber\\
&=M (\bar u u +\bar d d +\bar s s)^2.
\end{align}
Since $(\bar u u +\bar d d +\bar s s)$ is a SU(3) scalar, then the scalar that we form with the combination of Eq. (\ref{eq:M_matrix-2}) is
\begin{equation*}\label{eq:trace}
 {\rm Trace} [M (\bar u u +\bar d d +\bar s s)^2] = (\bar u u +\bar d d +\bar  s s)^3= {\rm Trace} [MMM].
\end{equation*}
Next we write the matrix $M$ in terms of the pseudoscalar mesons, taking into account the $\eta \eta'$ mixing \cite{bramon} and we obtain \cite{dani}
\begin{widetext}
\begin{equation}\label{eq:phimatrix}
M \to \phi \equiv \left(
           \begin{array}{ccc}
             \frac{1}{\sqrt{2}}\pi^0 + \frac{1}{\sqrt{3}}\eta + \frac{1}{\sqrt{6}}\eta' & \pi^+ & K^+ \\
             \pi^- & -\frac{1}{\sqrt{2}}\pi^0 + \frac{1}{\sqrt{3}}\eta + \frac{1}{\sqrt{6}}\eta' & K^0 \\
            K^- & \bar{K}^0 & -\frac{1}{\sqrt{3}}\eta + \sqrt{\frac{2}{3}}\eta' \\
           \end{array}
         \right).
\end{equation}
\end{widetext}
Then, the combination of three mesons that behaves as a SU(3) scalar is given by
\begin{equation}\label{eq:phiphiphi}
 {\rm SU(3)[scalar]} \equiv {\rm Trace} (\phi \phi \phi).
\end{equation}
By performing the algebra involved in Eq. (\ref{eq:phiphiphi}) and isolating the $\eta$ term we find the combination
\begin{equation}\label{eq:eta_term}
 C_1 :~~ \eta \left( \frac{6}{\sqrt{3}} \pi^+ \pi^-
+ \frac{3}{\sqrt{3}} \pi^0 \pi^0 + \frac{1}{3 \sqrt{3}} \eta \eta \right).
\end{equation}
Thus, when taking the structure of $V_1$ of Eq. (\ref{eq:V_structure}), apart from a $\eta$ in $P$-wave we shall have a $\pi^+ \pi^-$, $\pi^0 \pi^0$ or $\eta \eta$ produced in the primary step which will undergo final state interaction to produce a $\pi^+ \pi^-$. The $\eta$ will in principle interact with the pions but this would involve a $P$-wave, where the interaction is very weak and negligible in the energy region of interest to us \cite{Bernard:1991xb}. We shall explicitly take into account the $\pi \pi$ or $\eta \eta$ interaction in $S$-wave \cite{npa}, which will give rise to the $f_0(500)$, $f_0(980)$ resonances. We shall take into account the contribution of the $f_2(1270)$ empirically.
The $f_2(1270)$ appears within the chiral unitary approach as a bound state of $\rho \rho$ in $S$-wave \cite{Molinavec,gengvec} and decays into $\pi \pi$ in $D$-waves. This resonance gives a small contribution in the $\pi^+ \pi^-$ distribution in the region of the $f_0(500)$ and $f_0(980)$ that we are concerned about,
and we take it into account to allow for a proper comparison with the data.

Similarly, if we isolate one pion to carry the $P$-wave, taking for instance the term $V_2$ in Eq. (\ref{eq:V_structure}), then we find the combination
\begin{equation}\label{eq:pi1_term}
 C_2 :~~ \pi^+ \left( \frac{6}{\sqrt{3}} \pi^- \eta
+3 K^0 K^- \right)
\end{equation}
and equivalently the term with $V_3$ in Eq. (\ref{eq:V_structure}) comes with the combination
\begin{equation}\label{eq:pi2_term}
 C_3 :~~ \pi^- \left( \frac{6}{\sqrt{3}} \pi^+ \eta
+3 K^+ \bar K^0 \right).
\end{equation}

Once again, we shall now allow the $\pi \eta$ in each of these combinations to interact in $S$-wave, which will give rise to a big signal of the $a_0(980)$. Note that in the $C_2$ and $C_3$ combinations, the $\pi \eta$ interaction in $P$-wave is negligible, and since the $\pi \pi$ system is necessarily produced in $I=0$, then it can not interact in $P$-wave either (in fact there is no trace of $\rho$ production in the experiment).

In the present process, we shall have the combination of the three structures of Eq. (\ref{eq:V_structure}) and then the primary amplitude will be of the type
\begin{equation}\label{eq:t_amplitue}
 t= A~ \vec{\epsilon}_{\chi_{c1}} \cdot \vec{p}_{\eta} + B~ \vec{\epsilon}_{\chi_{c1}} \cdot \vec{p}_{\pi^{+}} + C~ \vec{\epsilon}_{\chi_{c1}} \cdot \vec{p}_{\pi^{-}},
\end{equation}
and the first thing to note is that there is no interference between these terms. Indeed, the crossed terms in $|t|^2$ after averaging over the polarization of the massive $\chi_{c1}$ state go as
\begin{align}\label{eq:cross_term}
& {\overline{ \sum}} 2 {\rm Re}(A B^*)\vec{\epsilon}_{\chi_{c1}} \cdot \vec{p}_{\eta}~\vec{\epsilon}_{\chi_{c1}} \cdot \vec{p}_{\pi^{+}} \nonumber \\
=& 2 {\rm Re}(A B^*) \frac{1}{3} \delta_{ij} p_{\eta i} p_{\pi^{+} j} = \frac{2}{3} {\rm Re}(A B^*)  \vec{p}_{\eta} \cdot \vec{p}_{\pi^{+}},
\end{align}
 which will vanish upon integration over angles in phase space. Thus, for $|t|^2$ we shall have the sum of the squares of each amplitude in Eq. (\ref{eq:t_amplitue}) which are described below.

\begin{figure}[tb]\centering
\includegraphics[scale=0.54]{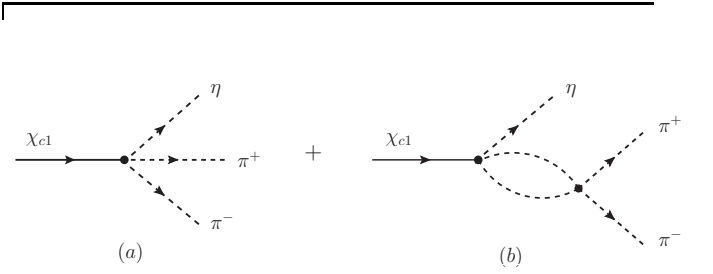}
\caption{Production of $\eta \pi^+ \pi^-$ through tree level (a) or rescattering (b) of $\pi^+ \pi^-$ pair.
\label{fig:FeynmanDiag3}}
\end{figure}

Next, we must take into account the final state interaction. For the process corresponding to $V_1$ of Eq. (\ref{eq:V_structure}) we can have
$\eta \pi^+ \pi^-$ in the final state by considering the $C_1$ combination of Eq. (\ref{eq:eta_term}) as depicted in Fig. \ref{fig:FeynmanDiag3}.
We will have
\begin{equation}\label{eq:t_eta}
t_{\eta}=\left( \vec{\epsilon}_{\chi_{c1}} \cdot \vec{p}_{\eta} \right) \tilde{t}_{\eta},
\end{equation}
with
\begin{equation}\label{eq:t_eta_tilde}
\tilde{t}_{\eta}=V_P \left( h_{\pi^+ \pi^-}+\sum_i h_i S_i G_i(M_{\rm inv}) t_{i,\pi^+ \pi^-} \right),
\end{equation}
where
\begin{equation}\label{eq:h_factor-1}
  h_{\pi^+ \pi^-}=\frac{6}{\sqrt{3}},~~~~~~ h_{\pi^0 \pi^0}=\frac{3}{\sqrt{3}},~~~~~~ h_{\eta \eta}=\frac{1}{3\sqrt{3}}
\end{equation}
are the weights of Eq. (\ref{eq:eta_term}) and $S_i$ are symmetry and combination factors for the identical particles,
\begin{equation}\label{eq:S_factor}
  S_{\pi^0 \pi^0}=2\times \frac{1}{2} ~~({\rm for~ two}~ \pi^0);~~~~~S_{\eta \eta}=3! \frac{1}{2} ~~({\rm ~for~ three} ~\eta).
\end{equation}
The functions $G_i$ and $t_{i, \pi^+ \pi^-}$ are the meson-meson loop functions and scattering amplitudes, which we take from Ref. \cite{npa} updated in Ref. \cite{liang,dai}.

\begin{figure}[tb]\centering
\includegraphics[scale=0.54]{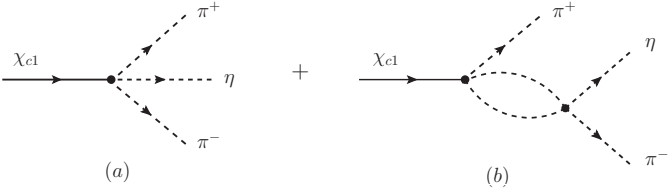}
\caption{Production of $\pi^+ \eta  \pi^-$ through tree level (a) and rescattering (b) of $\eta \pi^-$ pair.
\label{fig:FeynmanDiag4}}
\end{figure}

Similarly, corresponding to $V_2$ of Eq. (\ref{eq:V_structure}), we would have the mechanism depicted in Fig. \ref{fig:FeynmanDiag4}.
The amplitude corresponding to the diagrams of Fig. \ref{fig:FeynmanDiag4} is given by
\begin{equation}\label{eq:t_pi+}
t_{\pi^+}=\left( \vec{\epsilon}_{\chi_{c1}} \cdot \vec{p}_{\pi^+} \right) \tilde{t}_{\pi^+},
\end{equation}
with
\begin{equation}\label{eq:t_pi+_tilde}
\tilde{t}_{\pi^+}= V_P \left( h_{\pi^- \eta}+\sum_i h_i G_i(M_{\rm inv}) t_{i,\pi^- \eta} \right),
\end{equation}
and
\begin{equation}\label{eq:h_factor-2}
h_{\pi^- \eta}=\frac{6}{\sqrt{3}},~~~h_{K^0 K^-}=3.
\end{equation}
For the process associated to $V_3$ of Eq. (\ref{eq:V_structure}), we would have \footnote{The diagrams are similar to those of Fig. \ref{fig:FeynmanDiag4}.}
\begin{equation}\label{eq:t_pi-}
t_{\pi^-}=\left( \vec{\epsilon}_{\chi_{c1}} \cdot \vec{p}_{\pi^-} \right) \tilde{t}_{\pi^-},
\end{equation}
with
\begin{equation}\label{eq:t_pi-_tilde}
\tilde{t}_{\pi^-}=V_P \left( h_{\pi^+ \eta}+\sum_i h_i G_i(M_{\rm inv}) t_{i,\pi^+ \eta} \right),
\end{equation}
and
\begin{equation}\label{eq:h_factor-3}
h_{\pi^+ \eta}=\frac{6}{\sqrt{3}},~~~h_{\bar K^0 K^+}=3.
\end{equation}

As mentioned before, the interaction of the meson that comes with the $P$-wave with any of the other two, should proceed in $P$-wave, which is negligible for $\pi \eta$ and zero for $\pi \pi$ which have been created in $I=0$. This makes the interpretation of the signals particularly easy in this case, since they come from either the $\pi \pi$ or $\eta \pi$ interaction in $S$-wave.

The amplitudes for $\pi \pi, K\bar K, \pi\eta$ interaction are taken from Refs. \cite{liang,dai}, where only the neutral components are considered. Here we also need the charged components, which can easily be obtained using isospin symmetry and we find \cite{Xie:2015mzp}
\begin{align}\label{eq:t_structure}
&t_{K^0 K^-, \pi^- \eta} = \sqrt{2} t_{K^+ K^-, \pi^0 \eta}, \nonumber \\
&t_{K^+ \bar K^0, \pi^+ \eta} = \sqrt{2} t_{K^+ K^-, \pi^0 \eta}, \\
&t_{\pi^+ \eta, \pi^+ \eta} =  t_{\pi^- \eta, \pi^- \eta}=t_{\pi^0 \eta, \pi^0 \eta}. \nonumber
\end{align}

With all these ingredients we can write the differential mass distribution for $\pi^+ \pi^-$ as
\begin{equation}\label{eq:dGamma_pipi}
  \frac{d \Gamma}{d M_{\rm inv}(\pi \pi)}=\frac{1}{(2\pi)^3}\frac{1}{4M_{\chi_{c1}}^2}\frac{1}{3}p_{\eta}^2 p_{\eta} \tilde{p}_{\pi} \left| \tilde{t}_{\eta} \right|^2,
\end{equation}
where $p_{\eta}$ is the $\eta$ momentum in the $\chi_{c1}$ rest frame
\begin{equation}\label{eq:peta}
 p_{\eta}=\frac{\lambda^{1/2}(M_{\chi_{c1}}^2, m_{\eta}^2, M_{\rm inv}^2(\pi \pi))}{2M_{\chi_{c1}}},
\end{equation}
and $\tilde{p}_{\pi}$ is the pion momentum in the $\pi^+ \pi^-$ rest frame
\begin{equation}\label{eq:ppi_tilde}
 \tilde{p}_{\pi}=\frac{\lambda^{1/2}(M_{\rm inv}^2(\pi \pi), m_{\pi}^2, m_{\pi}^2)}{2M_{\rm inv}(\pi \pi)}.
\end{equation}

For the case of $\pi \eta$ invariant mass we would sum the contributions of $\pi^+ \eta$ and $\pi^- \eta$, which would give the same contribution, hence, the formula for $ \frac{d \Gamma}{d M_{\rm inv}(\pi \eta)}$, to be compared with experiment, will be
\begin{equation}\label{eq:dGamma_pieta}
  \frac{d \Gamma}{d M_{\rm inv}(\pi \eta)}=\frac{2}{(2\pi)^3}\frac{1}{4M_{\chi_{c1}}^2}\frac{1}{3}p_{\pi}^2 p_{\pi} \tilde{p}_{\eta} \left| \tilde{t}_{\pi^+} \right|^2,
\end{equation}
where now
\begin{equation}\label{eq:ppi}
 p_{\pi}=\frac{\lambda^{1/2}(M_{\chi_{c1}}^2, m_{\pi}^2, M_{\rm inv}^2(\pi \eta))}{2M_{\chi_{c1}}},
\end{equation}
\begin{equation}\label{eq:ppi_tilde2}
 \tilde{p}_{\eta}=\frac{\lambda^{1/2}(M_{\rm inv}^2(\pi \eta), m_{\pi}^2, m_{\eta}^2)}{2M_{\rm inv}(\pi \eta)}.
\end{equation}
The factor $V_P$ is the only unknown quantity in our approach, which provides a global normalization, and it is fitted to the data.
Note that the factors $A, B$ and $C$ in Eq. (\ref{eq:V_structure}) are absorbed in factor $V_P$.

\begin{figure}[tb]\centering
\includegraphics[scale=0.35]{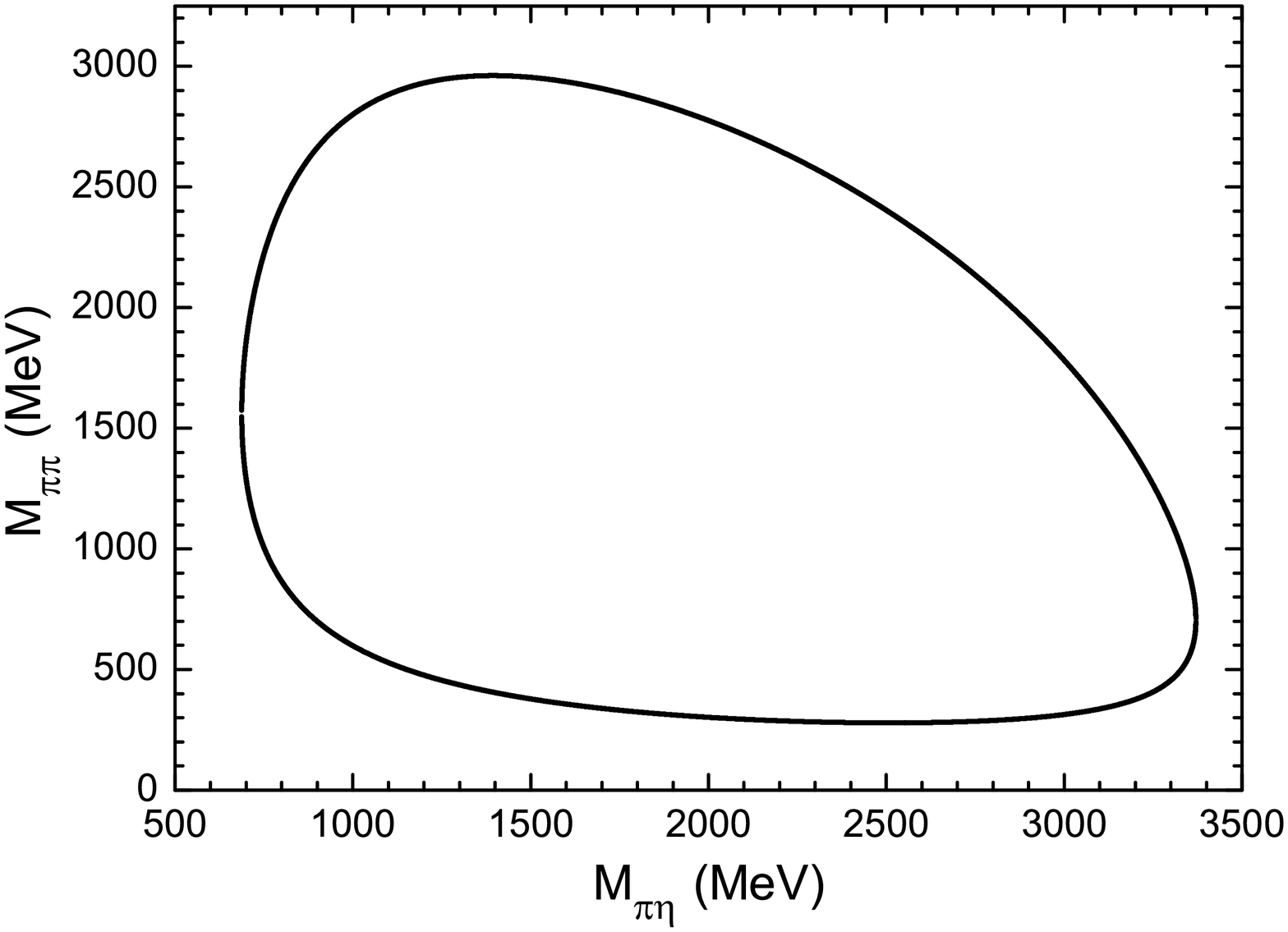}
\caption{Dalitz plot for the $\chi_{c1} \to \eta \pi^+ \pi^-$ decay.
\label{fig:DalitzPlot}}
\end{figure}

In principle we could have summed all the amplitudes and use the general $\frac{d^2 \Gamma}{d M_{\rm inv}(\pi \pi)~d M_{\rm inv}(\pi \eta)}$ formula, integrating over each of them to find the invariant mass distribution of the other pair. In practice, we find it unnecessary for the comparison of our results with data in the relevant region of invariant masses. The reason can be seen in the Dalitz plot that we show in Fig. \ref{fig:DalitzPlot}

If we consider $\frac{d \Gamma}{d M_{\rm inv}(\pi \eta)}$
 in the region of the $a_0(980)$, the $\pi \pi$ invariant mass has a range between 500 MeV and 2800 MeV. So its strength is divided over a large range of $\pi \pi$ invariant masses, providing a smooth background in the $\pi \pi$ mass distribution. We will take this into account empirically, following the analysis of Ref. \cite{kornicer}.

There is another element to consider. We have taken the $\pi \eta$ invariant mass from the interacting pair of Fig. \ref{fig:FeynmanDiag4}(b). Since, one is summing $\pi^+ \eta$, and $\pi^- \eta$ distributions, one would also have to account for the invariant mass distribution of the $\eta$ with the odd pion carrying the $P$-wave. Yet, it is easy to see where this mass distribution goes. Indeed, using the property
\begin{equation*}
 m^2_{12} + m^2_{13} +m^2_{23} = M^2_{\chi_{c1}} + m^2_{\pi}+m^2_{\pi}+m^2_{\eta},
\end{equation*}
taking $m_{23}=980 {\rm MeV}$ and $m_{12}\simeq 1200 {\rm MeV}$ (in the middle of the phase space allowed in the Dalitz plot), we find $m_{13} \simeq 2968 {\rm MeV}$, which is very far away from the region of the $a_0(980)$ and does not disturb the shape and strength of the $a_0(980)$.

Similarly, we will find a large contribution in the $\pi \pi$ invariant mass distribution from the $f_0(500)$ ($\sigma$ meson). Once again, by looking at the Dalitz plot, we see the strength is distributed in a region of $\pi \eta$ invariant masses from 1200 MeV to 3400 MeV, again a large region of invariant masses, but the most important for our discussion is that it does not contribute below the peak of the $a_0(980)$. Thus, the signal for the $a_0(980)$ is clean and easy to interpret, coming  basically from the $\pi \eta$ interaction.

\section{Results}
\begin{figure}[tb]\centering
\includegraphics[scale=0.37]{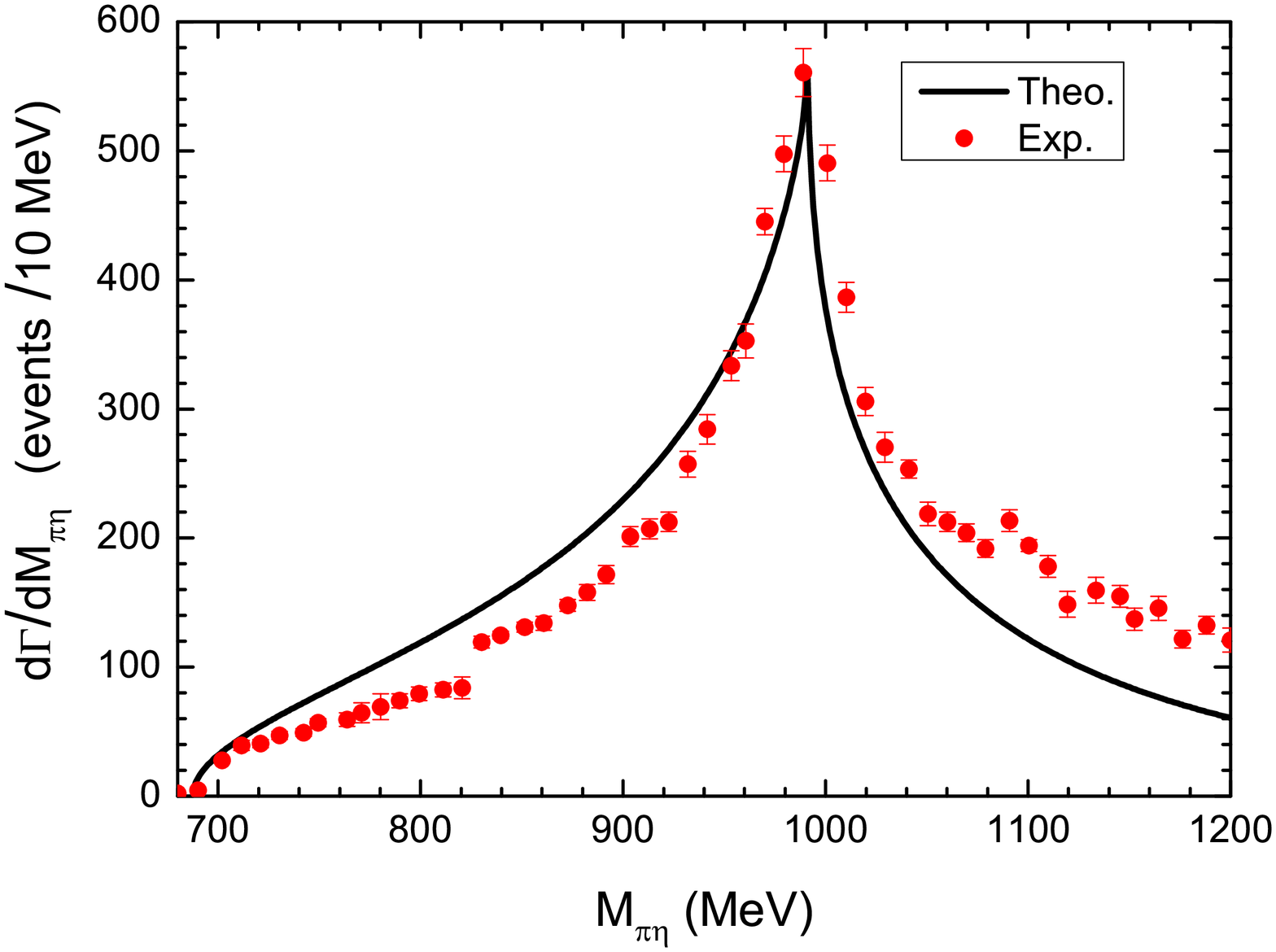}
\caption{$\pi \eta$ invariant mass distribution for the $\chi_{c1} \to \eta \pi^+ \pi^-$ decay. Preliminary BESIII data from Ref. \cite{kornicer}.
\label{fig:Mpieta}}
\end{figure}

In Fig. \ref{fig:Mpieta}, we show our results for the $\pi \eta$ invariant mass distribution. The parameter $V_P$ has been adjusted to the strength of the experimental preliminary data of BESIII at its peak \cite{kornicer}. As we can see, both the theory and the experiment show the typical huge cusp form of the $a_0(980)$. The agreement of our results with experiment is quite good, and some missing strength from 1100 MeV on can be clearly attributed to background from other sources. As we mentioned before, one does not see in the experiment much trace of a background below 1000 MeV.

\begin{figure}[tb]\centering
\includegraphics[scale=0.37]{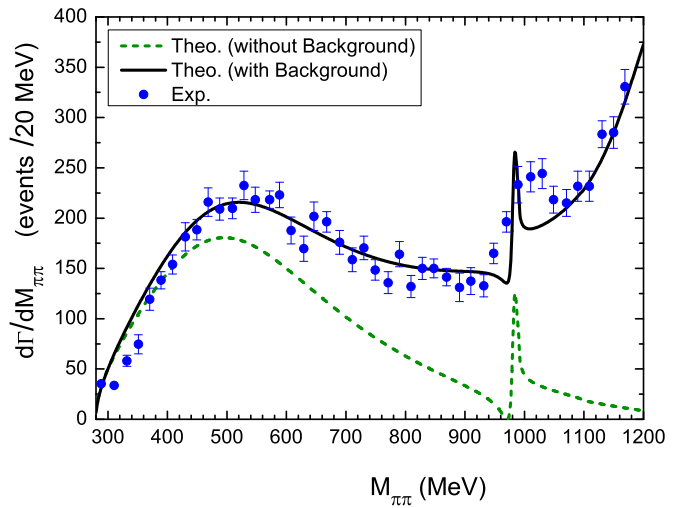}
\caption{$\pi \pi$ invariant mass distribution for the $\chi_{c1} \to \eta \pi^+ \pi^-$ decay. Preliminary BESIII data from Ref. \cite{kornicer}. The dashed line is the theoretical prediction. The solid line adds an empirical background(see text).
\label{fig:Mpipi}}
\end{figure}

In Fig. \ref{fig:Mpipi} we plot the invariant mass distribution for the $\pi^+ \pi^-$, using the same $V_P$ factor determined before. What we see is a relatively large strength for the production of the $f_0(500)$ and a small contribution from the $f_0(980)$. The experiment reflects both, a broad peak in the $f_0(500)$ region, and a rapid increase of the distribution in the region of $980 ~{\rm MeV}$. Our contribution of the $f_0(980)$ is rather sharp, while the experiment has a resolution of 20 MeV, and the raise of $\frac{d \Gamma}{d M_{\rm inv}(\pi \pi)}$ around 980 MeV is not so sharp. We should note that the strength of the $f_0(500)$ at its peak is about 110 events/10 MeV, compared to 560 events/10 MeV of the $a_0(980)$ at its peak. The signal for the $a_0(980)$ is thus quite big. Even integrating the strength of the $a_0(980)$ up to 1200 MeV and the one of the $f_0(980)$ up to 1000 MeV, we find a strength for the $a_0(980)$ almost 2.7 times bigger than that of the $f_0(980)$.

It is interesting to recall that the features of the $\pi \pi$ mass distribution are remarkably similar to those of the $J/\psi \to \omega \pi \pi$ reaction measured in Refs. \cite{Wu:2001vz,Augustin:1988ja}, which was studied along similar lines as here in Refs. \cite{Meissner:2000bc,Roca:2004uc}.

The features observed here are also similar to those observed in the
$\bar B^0 $ decay into $D^0$ and $f_0(500)$, $f_0(980)$ and $a_0(980)$
\cite{liangBD}, yet the relative strength of the structures found is
different, and is related to the weight of the different meson-meson
components prior to the final state interaction. The fact that one
describes all these reactions with this picture, and the chiral unitary
approach for the meson-meson interaction, offers support for the picture
of these resonances as dynamically generated from the meson-meson
interaction. Together with other reactions mentioned in the
Introduction, the support for this picture is, indeed, remarkable.

To facilitate the comparison with the data, we have added a background, very similar to the one of Ref. \cite{kornicer} coming from
the $a_0(980)$  peak, and which we have taken linear in the invariant
mass for simplicity. In addition, to account for the tail of  the
$f_2(1270)$, which shows up in Fig. 7 at high invariant masses, we have
taken a Breit Wigner shape, with physical mass and width, and adjusted
the strength to reproduce the data around 1100-1200 MeV. The agreement
with the $\pi \pi$ mass distribution is quite good, with some
discrepancy around 1000-1040 MeV. As mentioned above, the data shows a
fast raise around 980 MeV as our theory predicts, only that the
theoretical raise is sharper than experiment, where data are collected
in bins of 20 MeV.
On the other hand, the data shows a peak around 1040 MeV that the theory
cannot reproduce, even if we convolute the $f_0(980)$ signal with the
experimental resolution. The discrepancy is in two experimental points
and it would be worth checking whether this is just a fluctuation or a
genuine peak. We should note that in Ref. \cite{Adams:2011sq}, the data, with admittedly
smaller statistics, one does not see a structure around 1000-1060 MeV
like in Ref. \cite{kornicer}.

  In any case, the data of Ref. \cite{kornicer} is also telling us that the strength
of the $f_0(980)$ is far smaller than the one of the $f_0(500)$, as the
theory predicts. It would be interesting to see what comes out from the
final analysis of  Ref. \cite{kornicer}, which motivated our work. A partial wave
analysis can separate the contribution of the different structures,
allowing for a more quantitative comparison with our results.

\section{Conclusions}
We have made a study of the $\chi_{c1} \to \eta \pi^+ \pi^-$ reaction,
looking at the $\pi^+ \pi^-$ and $\eta \pi$ invariant mass
distributions. We have used a simple picture to combine the mesons to
give a singlet of SU(3), which corresponds to the $c \bar c$ nature of
the $\chi_{c1}$. This gives us the relative weights of three mesons at a
primary production step, which can revert into the $\eta \pi^+ \pi^-$ in
the final state upon interaction of pairs of mesons in coupled channels.  We have used the
chiral unitary approach to describe this interaction, which generates
the $f_0(500)$, $f_0(980)$ and $a_0(980)$ scalar mesons. The interesting
feature of the approach is that, up to a global normalization constant,
we are able to construct the $\pi \pi$ and $\eta \pi$ invariant mass
distributions and compare with the experimental data available. We
observed a prominent signal of the $a_0(980)$ production with a relative
strength to the other two resonances much bigger than in other reactions
studied previously. We also observed a clear signal for $f_0(500)$
production in the $\pi^+ \pi^-$ mass distribution, and also a clear signal for $f_0(980)$ production, but
with much smaller strength. The
agreement with experiment is quite good in the two invariant mass
distributions up to about 1040 MeV, once a background borrowed from the
experiment is implemented in the $\pi \pi$ distribution. We also
justified that no background for the $\eta \pi$ distribution was needed
in that energy range.

  We noted some small discrepancy with the data around 1040 MeV that
could be given extra attention in the final analysis of the work of Ref.
\cite{kornicer}.

   The agreement found in general lines for the shapes and strengths of
the  $f_0(500)$, $f_0(980)$ and $a_0(980)$ excitation in this reaction
adds to the long list of reactions that give support to these resonances
as being dynamically generated from the interaction of pseudoscalar mesons.

\section*{Acknowledgments}
One of us, E. O. wishes to acknowledge support from the Chinese Academy
of Science in the Program of Visiting Professorship for Senior International Scientists (Grant No. 2013T2J0012).
This work is partly supported by the
National Natural Science Foundation of China under Grants
No. 11565007, No. 11547307 and No. 11475227. It is also supported by the Youth Innova-
tion Promotion Association CAS (No. 2016367).
This work is also partly supported by the Spanish Ministerio
de Economia y Competitividad and European FEDER funds
under the contract number FIS2011-28853-C02-01, FIS2011-
28853-C02-02, FIS2014-57026-REDT, FIS2014-51948-C2-
1-P, and FIS2014-51948-C2-2-P, and the Generalitat Valenciana
in the program Prometeo II-2014/068.

\clearpage

\bibliographystyle{plain}

\end{document}